\documentclass[12pt]{article} 
\usepackage{graphicx,amsmath} 

\def\fun#1#2{\lower3.6pt\vbox{\baselineskip0pt\lineskip.9pt
  \ialign{$\mathsurround=0pt#1\hfil##\hfil$\crcr#2\crcr\sim\crcr}}}

\newcommand{\Li}{\mbox{Li}_2}
\newcommand{\dd}{\mbox{d}}

\newcommand{\ii}{\mbox{i}}
\newcommand{\vecc}[1]{\mbox{\boldmath $#1$}}

\newcommand{\be}{\begin{equation}}
\newcommand{\ee}{\end{equation}}
\newcommand{\ba}{\begin{eqnarray}}
\newcommand{\ea}{\end{eqnarray}}
\newcommand{\bg}{\begin{gather}}
\newcommand{\eg}{\end{gather}}

\title{ (Quasi)elastic electron-muon large-angle scattering
to a two-loop approximation: eikonal type contributions I.
}

\author{
V.V.~Bytev$^\sharp$\footnote{supported in part by RFBR 01-02-17437 and INTAS 00366 grants.},
E.A.~Kuraev$^\sharp$\footnote{supported in part by RFBR 01-02-17437 and INTAS 00366 grants.},
B.G.~Shaikhatdenov$^{\sharp\natural}$\footnote{supported by Conacyt (M\'exico).}
\vspace{4mm} \\
$^\sharp${\small\sl Joint Institute for Nuclear Research, Dubna 141980 Russia}\\
$^\natural${\small\sl Departamento de F\'{\i}sica, Cinvestav del IPN,
07000 M\'exico D.F.} }

\date{}
\begin{document}
\maketitle

\begin{abstract}
Part of eikonal type contributions to $e\mu$ large-angle high-energy
scattering cross section is considered in a quasi-elastic experimental
set-up. Apart from virtual corrections we examine inelastic processes
with emission of one and two soft real photons as well as soft lepton
and pion pairs.
Virtual photon contributions are given within a logarithmic accuracy.
Box type Feynman amplitudes with leptonic and hadronic vacuum polarization
insertion are considered explicitly as well as double box ones.
Wherever appropriate analytic expressions obtained are compared
with those predicted by a structure function approach.
\end{abstract}

\section{Introduction}
This paper is a follow-up of the previous one devoted to the evaluation
of a gauge-invariant set of contributions to the cross section
of a quasielastic large-angle $e\mu$ scattering at the second order
in the coupling constant of perturbation theory (PT)~\cite{BKS02}.


The need for the evalution of the radiative corrections (RC) at two-loop
order is dictated by the experimental data on observables
for a collider calibration process of electron-positron scattering
that is reached nowadays an impressive level of accuracy.
Inspired by this we consider the determination of the second order RC
to the cross section of Bhabha scattering to be our ultimate goal.
At the same time, since the task of two-loop calculus is rather involved
it appears to be
more easier to look first at electron-muon scattering despite different
masses of interacting particles.
Besides the latter process is important in itself for that it forms
a background to the rare processes in particular those violating
lepton number (for more details see~\cite{BKS02} and references therein).
Therefore improving theoretical predictions
on its observables could place more stringent bounds on the physics
beyond the Standard Model.

The aim of this investigation is to calculate the next-to-leading order
contributions to the electron-muon large-angle high-energy cross section
\be
\nonumber
e^-(p_1)+\mu^-(p_2)\to e^-(p'_1) + \mu^-(p'_2),
\ee
in a quasielastic experimental set-up
\be \label{123}
\frac{2\varepsilon- \varepsilon'_1 - \varepsilon'_2}{2\varepsilon}
=\frac{\Delta\varepsilon }{\varepsilon}\equiv\Delta\ll 1,
\qquad \Delta \varepsilon\gg m_\mu(m_\pi)\,,
\ee
where $\varepsilon, \varepsilon'_1,\varepsilon'_2$ are the energies
of initial and scattered leptons in the center-of-mass reference frame
and the Mandelstam variables are much larger than any involved
particle's mass squared. The quantity $\Delta\varepsilon$
indicates an energy resolution of detectors supposed to track final particles.
The cross section possesses in the leading logarithmic approximation (LLA)
the form of the Drell-Yan process's cross section~\cite{KF85},
\be\label{eq:DY}
\dd\sigma(s,t)=\int\prod\limits_{i=1}^4\dd x_i
{\cal D}(x_i,\rho_t)\,\dd\sigma_0(sx_1x_2,tx_1x_3)\left(1+\frac{\alpha}{\pi}K\right)\,,
\ee
where
\be
\nonumber
\rho_t=\ln\frac{-t}{m_em_\mu},\quad t=(p_1-p'_1)^2,\quad s=(p_1+p_2)^2,
\quad u=(p_1-p'_2)^2\,.
\ee
In the above expression the quantities ${\cal D}(x_i,\rho_t)$ are
the non-singlet structure functions which obey renormalization group (RG)
evolution equations. Their expansion in LLA
($(\alpha/\pi)\ll 1,\ (\alpha/\pi)\rho_t\sim 1$) can be written as,
\be\label{1232}
{\cal D}(x,\rho_t)=\delta(1-x)+\sum\limits_{n=1}^{\infty}\frac{1}{n!}
\left(\frac{\alpha\rho_t}{2\pi}\right)^n{\cal P}^{(n)}(x)\,.
\ee
In a quasielastic set-up it is appropriate to use only $\delta$-part
of the splitting function ${\cal P}^{(n)}(x)$  denoted by
${\cal P}_\Delta^{(n)}(x)$,
\ba
{\cal P}^{(n)}(x)=\int\limits_x^1\frac{\dd y}{y}{\cal P}^{(1)}(y){\cal P}^{(n-1)}
\left(\frac{x}{y}\right),\!\!\!\!&&\!\!\! n\geq 2, \nonumber \\
{\cal P}^{(1)}(x)=\left(\frac{1+x^2}{1-x}\right)_+&=&\lim\limits_{\Delta\to 0}
\bigl[ {\cal P}^{(1)}_\Delta(x) + {\cal P}^{(1)}_\theta(x)\bigr],
\\ \nonumber
{\cal P}^{(1)}_\Delta(x)={\cal P}^{(1)}_\Delta \delta(1-x),
\qquad {\cal P}^{(1)}_\Delta&=&2\ln\Delta + \frac{3}{2},
\qquad P^{(1)}_\theta(x)=\frac{1+x^2}{1-x}\Theta(1-x-\Delta).
\ea
Then the structure function gets the following form,
\ba
{\cal D}(x,\rho_t)=\delta(1-x)\biggl[1+\sum_{n=1}^{\infty}\frac{1}{n!}
\left(\frac{\alpha\rho_t}{2\pi}\right)^n{\cal P}_\Delta^{(n)}\biggr].
\ea
As the structure function approach outlined is capable of providing only
the leading log corrections we need to explicitly calculate
the so called $K$-factor entering Eq.(\ref{eq:DY}) in a one
and two-loop approximations.

Broadly speaking the RC to the differential cross section
in the mass regularization scheme adopted are of two types.
The first one is that arising from virtual photon emission
up to second order of PT,
which requires the calculations of, amongst others, the real two-loop
Feynman amplitudes.
Those suffer from infrared divergences, which are regularized
by assigning a negligibly small mass $\lambda$ to the photon,
to be set to vanish at the end of the calculations.
The second type of contributions comes from the emission of soft real photons
and charged particle pairs.

The general structure of the correction to the cross section can be presented
as a sum of three types: vertex, eikonal and decorated box type.
Each of them contains virtual and real soft photon contributions and is free of
infrared divergences. They are all do not violate the structure of the
leading log correction predicted on the base of RG ideas
though the contributions of individual diagrams contain up to
fourth power of large logarithm $\rho_t$ at two-loop order.
In this regard we remind that in our previous paper~\cite{BKS02} it was
shown that the vertex contributions do already provide a result consistent
with RG approach.
\begin{figure}[htb!]
\begin{center}
\fbox{\includegraphics[height=2.7cm]{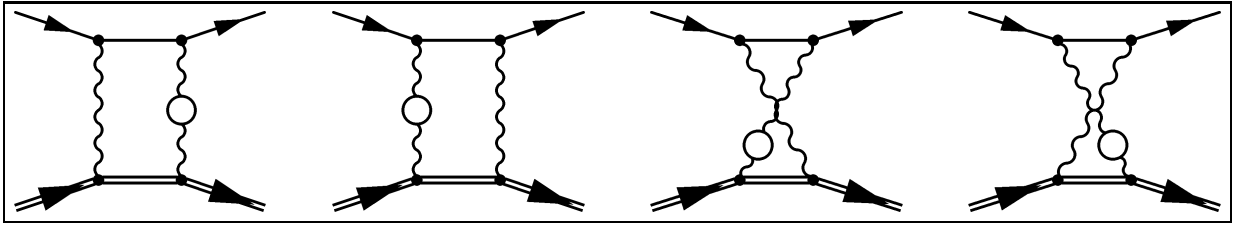}}
\end{center}
\caption{Box type graphs with a vacuum polarization insertion.}
\label{fig1}
\end{figure}
As the first order RC coming from
box type diagrams were given in our previous work devoted to the evaluation
of the vertex type contributions~\cite{BKS02} here we concentrate
on the investigation of some eikonal box type diagrams
at the second order of PT.
In the case of elastic process they correspond to graphs
with one, two (box diagram) and three (double box diagram)
virtual photon mediated between interacting leptons.
\begin{figure}[htb!]
\begin{center}
\fbox{\includegraphics[height=3.0cm]{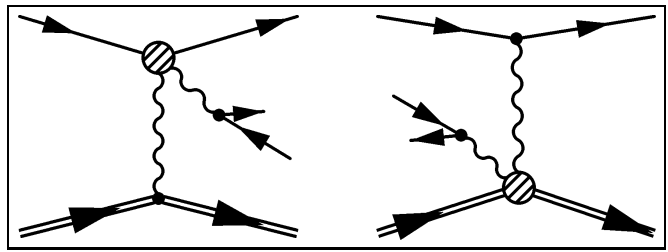}}
\end{center}
\caption{A soft lepton and pion pair production.}
\label{fig2}
\end{figure}
Besides we have to take into account
box type graphs with a vacuum polarization insertion into the
Green function of either of virtual exchange photons (see Fig.~\ref{fig1}).
A single soft photon approximation must be applied to the
one-loop corrected Feynman amplitudes in order to get another
set of contributions.
And finally at this order an emission of two soft photons
(pairs of charged particles) has to be as well taken into account.
\begin{figure}[htb!]
\begin{center}
\fbox{\includegraphics[height=3.3cm]{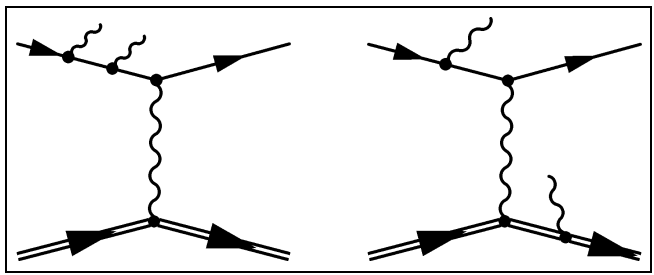}}
\end{center}
\caption{Sample diagrams pertinent to the double soft photon emission.}
\label{fig3}
\end{figure}
Let's briefly mention content of the paper.
In Sec.~2 we consider the vacuum polarization effects in box type
Feynman amplitudes with lepton
$(\mu\bar{\mu},e\bar e)$ and pion $(\pi^-\pi^+)$ pairs running a loop.
Also in this section we deal with a corresponding contribution coming from
a soft lepton pair as well as a soft charged pion pair production
along with the one soft photon emission (see Fig.\ref{fig2})
associated with the one-loop virtual exchange photon self-energy amplitudes.
In Sec.~3 the results of evaluation of the corrections
corresponding to a single and double soft photon emission (see Fig.~\ref{fig3})
as well as to a square of box type diagrams are presented followed
by the short concluding remarks.

In App.~A we
present a set of scalar integrals for the box type diagrams
with a vacuum polarization insertion.
In App.~B one can find some details of the derivation of RC coming from
the squared box type diagrams. There we as well give all the
integrals encountered during evaluation.

\section{Box type diagrams with a vacuum polarization insertion}
Vacuum polarization (VP) effects in the box type Feynman amplitudes can be
taken into account by replacing one of the photon propagators
by the VP insertion (see~\cite{SwBMR}). In the case when leptons with mass
$\mu$ run a loop it looks,
\ba
\frac{1}{k^2}\to \frac{\alpha}{3\pi}\int\limits_0^1\frac{\phi(v)\dd v}
{(1-v^2)(k^2-M^2(v))},\quad
M^2=\frac{4\mu^2}{1-v^2},\quad \phi(v)=2-(1-v^2)(2-v^2)\,,
\ea
and for a pion-antipion pair in the loop it reads,
\ba
\frac{1}{k^2}\to \frac{\alpha}{3\pi}\int\limits_{4m_{\pi}^2}^\infty
\frac{\dd M^2}{M^2}\frac{{\cal R}(M^2)}{k^2-M^2},
\qquad {\cal R}(M^2)=\frac{\sigma^{e\bar e\to hadr}(M^2)}
{\sigma^{e\bar e\to\mu\bar\mu}}\,.
\ea
Here the quantity $M$ stands for an invariant mass of hadronic jet
produced in a single-photon annihilation of a lepton pair
and the quantity
${\cal R}(M^2)$ is the known experimental input ratio~\cite{EE95}.
Then for the matrix element squared we obtain:
\ba\label{1235}
\delta|{\cal M}|^2_{vp(lept)}=\frac{2^8\alpha^4}{3t}\int\limits_0^1
\dd v\frac{\phi(v)}{1-v^2}\bigl[S(s,t,M^2)-S(u,t,M^2)\bigr]\,,
\ea
for VP induced by leptons, and
\ba\label{1236}
\delta|{\cal M}|^2_{vp(hadr)}=\frac{2^8\alpha^4}{3t}
\int\limits_{4m_\pi^2}^\infty\frac{\dd M^2}{M^2}{\cal R}(M^2)
\bigl[S(s,t,M^2)-S(u,t,M^2)\bigr]\,,
\ea
for the hadronic VP contribution.

The quantity $S(s,t,M^2)$ is universal irrespective of the virtual pair
running a self-energy loop and reads as follows,
\be
S(s,t,M^2)=\int\frac{\dd^4 k}{\ii\pi^2}
\frac{{\mathrm{Tr}}(e)\cdot{\mathrm{Tr}}(\mu)}{(1)(2)(3)(4)},
\ee
with
\ba\label{eq1}
(1)=k^2-2kp_1,\quad (2)=k^2+2kp_2,&& (3)=k^2-2kq+\tilde{t},
\quad (4)=k^2-\lambda^2, \nonumber \\
{\mathrm Tr}(e)=\frac{1}{4}Sp\{p_1\gamma_\mu p_1^{'}\gamma_\nu(p_1-k)
\gamma_\lambda\},
&&{\mathrm Tr}(\mu)=\frac{1}{4}Sp\{p_2\gamma_\mu p_2^{'}\gamma_\nu(p_2+k)
\gamma_\lambda\}, \\ \nonumber
p_1^2=m^2_e,\quad p_2^2=m^2_\mu,&& \tilde{t}=t-M^2,\quad q=p_1-p'_1.
\ea
Using a set of scalar, vector and tensor box type integrals
given in App.~A the quantity $S(s,t,M^2)$ can be expressed via a few
basic integrals,
\ba
S(s,t,M^2)&=&u\biggl[\ln\frac{s}{-\tilde{t}}
+\frac{M^2}{t}\ln\frac{-\tilde{t}}{M^2}\biggr]
-\left(s(s-u)+\frac{t\tilde{t}}{2}\right)\bigl[I_{134}+I_{234}\bigr]
\\ \nonumber
&+&s(s^2+u^2)I
+s\left(u+\frac{\tilde{t}}{2}\right)\bigl[-I_{123}-I_{124}
+\tilde{t}I\bigr]\,,
\ea
where
\ba
I_{ijl}=\int\frac{\dd^4k}{\ii\pi^2}\frac{1}{(i)(j)(l)},\qquad
I=\int\frac{\dd^4k}{\ii\pi^2}\frac{1}{(1)(2)(3)(4)}.
\ea
Performing loop-momentum integration and neglecting terms
of order $m_\mu^2/(-t)\ll 1$ we find in the limit of large invariant variables,
\ba\label{1233}
\biggl[S(s,t,M^2)&-&S(u,t,M^2)\biggr]\bigg|_{|t|\gg M^2}=\frac{s^2+u^2}{t}L_{us}
(\rho_m-2\rho_t-\rho_\lambda) \nonumber \\
&+&(u-s)\left(\frac{1}{2}\pi^2+\rho^2_m-\frac{1}{2}L_{st}^2-\frac{1}{2}L_{ut}^2
+\ln^2\frac{m_\mu}{m_e}\right) + uL_{st}-sL_{ut}\,, \\ \nonumber
\rho_m=\ln\frac{M^2}{m_em_\mu},&&
\rho_\lambda=\ln\frac{m_em_\mu}{\lambda^2},
\quad L_{st}=\ln\frac{s}{-t},\quad L_{ut}=\ln\frac{u}{t},
\quad L_{us}=\ln\frac{-u}{s}\,.
\ea
In the opposite limit the answer is found to be,
\ba\label{1234}
\biggl[S(s,t,M^2)&-&S(u,t,M^2)\biggr]\bigg|_{M^2\gg |t|}=\frac{1}{M^2}\biggl[
\frac{s^2+u^2}{2}L_{us}(\rho_s+\rho_u+2\rho_\lambda) \\ \nonumber
&+&\frac{3}{2}(u^2\rho_s-s^2\rho_u)
+t^2L_{us}+t(u-s)\left(\frac{3}{2}\rho_m+\frac{7}{4}\right)+\frac{s^2+u^2}{2}\pi^2\biggr].
\ea
Further integration in the case of leptonic VP
with the mass $M^2=4\mu^2/(1-v^2)$ (both cases $\mu=m_e, m_\mu$
are taken into account) leads within a logarithmic accuracy to the
following expression,
\ba\label{16}
\frac{\dd\sigma_{vp}^{box}}{\dd\sigma_0}=\frac{2\alpha^2}{3\pi^2}\rho_t\biggl\{
2L_{su}\left(\frac{3}{2}\rho_t + \rho_\lambda-\frac{10}{3}\right)
-\frac{s^2-u^2}{s^2+u^2}(L_{st}^2+L_{ut}^2-2\pi^2) \\ \nonumber
+\frac{2t}{s^2+u^2}[(t-s)L_{st}-(t-u)L_{ut}]
\biggr\}.
\ea
To finalize this result we have to remove infrared divergences. To this end
an interference between the soft photon emission tree level amplitudes
and those bearing a leptonic VP insertion has to be taken into account
yielding,
\ba\label{17}
\frac{\dd\sigma_{vp}^{\gamma}}{\dd\sigma_0}=-\frac{4\alpha^2}{3\pi^2}
\left(\rho_t-\frac{5}{3}\right)\biggl[
(2\ln\Delta+\rho_\lambda)L_{su}+\rho_tL_{su}
-\frac{1}{2}(L^2_{ut} - L^2_{st}) - \Li\biggl(\frac{1-c}{2}\biggr)\biggr],
\ea
with $\Delta$ given in Eq.~(\ref{123}), $c=\cos\widehat{p_1,p}'_1$ being
a cosine of the scatter angle in cms and the dilogarithm function
defined as usual,
\be
\nonumber
\Li(x)=-\int\limits_0^x\frac{\ln(1-t)}{t}\dd t.
\ee
Furthermore
we have to consider the contribution coming from a soft lepton pair production
with a total pair energy not exceeding $\Delta\varepsilon$
($2\mu\ll\Delta\varepsilon\ll\varepsilon$). This could be read off for example
from Ref.~\cite{AKMT97},
\ba\label{18}
\frac{\dd\sigma_{sp}}{\dd\sigma_0}= - \frac{2\alpha^2}{3\pi^2}\rho_t\biggl\{
L_{su}\left[\rho_t + L_{st} + L_{ut}
+2\left(2\ln\Delta-\frac{5}{3}\right)\right]
-2\Li\left(\frac{1-c}{2}\right)\biggr\}.
\ea
Then the ultimate result with logarithmic accuracy for a total correction given by leptonic VP
and a soft $e\bar e, \mu\bar\mu$ pair production is brought to the form
(see Eqs.~(\ref{16},\ref{17},\ref{18})),
\ba\label{1237}
\frac{\dd\sigma_{vp+sp}}{\dd\sigma_0}&=&\frac{2\alpha^2}{3\pi^2}\rho_t\biggl[
2(L_{ut}^2-L_{st}^2) + 8L_{us}\ln\Delta
- \frac{s^2-u^2}{s^2+u^2}(L_{ut}^2+L_{st}^2-2\pi^2) \nonumber \\
&+& \frac{2t}{s^2+u^2}(tL_{su}-sL_{st}+uL_{ut})
+ 4\Li\biggl(\frac{1-c}{2}\biggr) \biggr].
\ea
This expression is seen to contain only next-to-leading term
(of order $\alpha^2\rho_t$) and is free of infrared divergences.

Let's now turn to a soft pion pair production
with a total pair energy staying below $\Delta\varepsilon$ and invariant
mass squared $M^2$ bounded as follows,
\be
\nonumber
4m_\pi^2\ll M^2<(\Delta\varepsilon)^2\ll\varepsilon^2=s/4.
\ee
The corresponding contribution to the differential cross section
arises from the interference of ``up-down" pair production, that is
of pairs created by the virtual photons emitted off an electron and a muon lines,
\ba\label{eq:20}
\frac{\dd\sigma}{\dd M^2\dd\sigma_0}\bigg|_{\pi^\pm}
&=&2\biggl(\frac{4\pi\alpha}{M^2}\biggr)^2
\frac{\dd^4 q}{M^2}\int\frac{\dd^3{\vecc q}_+\dd^3{\vecc q}_-}
{2\varepsilon_+2\varepsilon_-}\delta^4(q_++q_--q) \\ \nonumber
&\times&\biggl(\frac{Qp'_1}{qp'_1}-\frac{Qp_1}{qp_1}\biggr)
\biggl(\frac{Qp'_2}{qp'_2}-\frac{Qp_2}{qp_2}\biggr),
\qquad q^2=M^2,\quad Q=q_+-q_-.
\ea
Carry first out an invariant pion pair phase space integration,
\be
\nonumber
\int\frac{\dd^3{\vecc q}_+\dd^3{\vecc q}_-}{2\varepsilon_+2\varepsilon_-}
\delta^4(q_++q_--q)Q_\mu Q_\nu=\frac{1}{3}\frac{\pi\beta}{2}
\biggl(g_{\mu\nu}-\frac{q_\mu q_\nu}{q^2}\biggr)Q^2,
\quad \beta=\sqrt{1-\frac{4m_\pi^2}{q^2}}.
\ee
Upon rearranging the phase volume,
\be\nonumber
\int\frac{\dd^4q}{\dd q^2}=\frac{1}{2}
\int\limits_{\sqrt{q^2}}^{\Delta\varepsilon}\dd q_0\sqrt{q_0^2-q^2}
\int\dd\Omega_q\,,
\ee
the rhs of Eq.~(\ref{eq:20}) can be recast to take the following form,
\be
\nonumber
\frac{\alpha^2}{3\pi^2}\int\limits_{M}^{\Delta\varepsilon}\dd q_0
\sqrt{q_0^2-M^2}
\int\frac{\dd\Omega_q}{4\pi}\biggl(\frac{p_1p_2}{p_1q\cdot p_2q}
-\frac{p_1p_2^{'}}{p_1q\cdot p'_2q}\biggr)
=\frac{\alpha^2}{3\pi^2}\biggl(L_{su}\ln\frac{2\Delta\varepsilon}{M}
+\mathcal{O}(1)\biggr).
\ee
The final result is then appears to be,
\ba\label{1238}
\frac{\dd\sigma}{\dd M^2\dd\sigma_0}\biggr|_{\pi^\pm}
=\frac{\alpha^2}{6\pi^2M^2}\bigl[ L_{su}(\rho_t-\rho_m) + {\cal O}(1)\bigr]\,.
\ea
Obviously the contribution coming from the box type diagrams
with hadronic VP cannot be obtained in analytical form
due to presence of the quantity ${\cal R}(M^2)$.

\section{Squared box and corresponding soft photon corrections}
An ``up-down" interference of a soft photon emission off an electron
and a muon lines can be evaluated by making use of the following expression,
\ba
I_{p_A p_B}&=&\frac{1}{4\pi}\int\frac{\dd^3{\mathbf k}}{\omega}
\frac{p_Ap_B}{p_Ak\cdot p_Bk}\bigg|_{\omega<\Delta\varepsilon}
\\ \nonumber
&=&\left(\ln\Delta+\frac{1}{2}\rho_\lambda\right)L_{AB}
+\frac{1}{4}\left(L_{AB}^2-\ln^2\frac{m_\mu}{m_e}\right)
-\frac{\pi^2}{6}+\frac{1}{2}\Li\biggl(\frac{1+c}{2}\biggr),
\ea
where
\ba
L_{AB}=\ln\left(\frac{2p_Ap_B}{m_em_\mu}\right),
\quad p_Ap_B=\varepsilon^2(1-c), \quad p_A^2=m_e^2,\quad p_B^2=m_\mu^2,
\quad \varepsilon_A=\varepsilon_B\equiv\varepsilon\,, \nonumber
\ea
and the quantity $\omega$ is a soft photon energy.
Using known results for the interference of the Born and box type
elastic amplitudes (see App.~B) we obtain for a single soft photon
emission contribution (in a soft photon approximation),
\ba\label{1239}
\frac{\dd\sigma_{\mathbf{box}}^\gamma}{\dd\sigma_0}=\biggl(\frac{\alpha}{\pi}\biggr)^2
\biggl[2L_{su}(\rho_t+\rho_\lambda)+\frac{t^2}{s^2+u^2}
\left(\frac{u}{t}L_{st}-\frac{s}{t}L_{ut}
+\frac{s-u}{2t}(\pi^2+L_{ut}^2+L_{st}^2)\right)\biggr] \\ \nonumber
\times\biggl[-L_{su}\rho_t+\frac{1}{2}(L_{ut}^2-L_{st}^2)
-2L_{su}\left(\ln\Delta+\frac{1}{2}\rho_\lambda\right)
+\Li\biggl(\frac{1-c}{2}\biggr)\biggr].
\ea
In the case of two soft photon emission
with a total energy again not exceeding $\Delta\varepsilon$ we have,
\ba\label{1230}
\frac{\dd\sigma^{\gamma\gamma}}{\dd\sigma_0}&=&
\left(\frac{2\alpha}{\pi}\right)^2
\Biggl\{\biggl[\frac{1}{2}\rho_tL_{su}+\frac{1}{4}(L_{st}^2-L_{ut}^2)
 \\ \nonumber
&+&L_{su}\left(\ln\Delta+\frac{1}{2}\rho_\lambda\right)
-\frac{1}{2}\Li\biggl(\frac{1-c}{2}\biggr)
\biggr]^2 - \frac{\pi^2}{6}L_{su}^2\Biggr\}.
\ea
And finally from the evaluation of the squared box type graphs given in App.~B
we infer the logarithmic contribution which is written as follows,
\ba\label{124}
\frac{\dd\sigma_{BB}}{\dd\sigma_0}=\frac{\alpha^2}{\pi^2}\frac{t^2}{s^2+u^2}\rho_t
\bigl[A\rho_t+B\bigr],
\ea
where the coefficients read,
\begin{eqnarray*}
A&=&2\frac{s^2+u^2}{t^2}(L_{us}^2+\pi^2), \\
B&=&4\frac{s^2+u^2}{t^2}(L_{us}^2+\pi^2)\rho_\lambda + 2L_{us}
\left(\frac{s}{t}L_{ut}-\frac{u}{t}L_{st}\right) \\
&+&\frac{s-u}{t}\left[\pi^2(2L_{st} - L_{us})
-L_{us}(L_{ut}^2+L_{st}^2)\right] + \frac{8u}{t}\pi^2.
\end{eqnarray*}

\section{Summary}
This paper is devoted to the determination of a part
of second order RC to the cross section of the process
of large-angle quasi-elastic $e\mu$ scattering, namely those corresponding
to eikonal box type diagrams.
In the case of the box type diagrams with a vacuum polarization insertion
we obtain the formul\ae{} given in Eqs.~(\ref{1233},\ref{1234},\ref{1238})
where it is seen that the contributions coming from the interference
between a tree level diagram and those (bearing a vacuum polarization insertion)
with straight and crossed ``legs" become in fact equal
when one exchanges $s\leftrightarrow u$ (with accuracy up to terms $\sim\pi^2$)
and alternates an overall sign of the contribution. This is
indeed a manifestation of the well-known symmetry relation
between amplitudes corresponding to different channels of a given reaction.

The main results of this work are given in analytic form in the logarithmic
approximation but the intermediate ones are presented to a power accuracy
thus allowing for at least numeric evaluation of the impact of subleading
terms on the overall value of the corrections. For example, in Sec.~2 we obtain
two limiting cases of the contribution due to leptonic VP,
for a small (Eq.~(\ref{1233})) and a large (Eq.~(\ref{1234})) lepton pair
invariant mass $M$ with constant accuracy

As a consistency check of the calculation, the auxiliary infrared
parameter $\lambda$ is expected to completely cancel out of the final results.
Therefore within a gauge invariant set of amplitudes considered in Sec.~2
we show that by integrating over the integration variable $v$
and then by adding the contribution given by a soft lepton pair production
one indeed achieves a result free of infrared divergences (Eq.~(\ref{1237})).
The structure of this correction is in agreement with RG predictions
and does not contain large logarithms of higher than second powers.
But the same cannot be done for the contributions calculated in Sec.~3 since
a consideration carried out there is in fact incomplete.
Also we give the expression for a cross section of a soft pion pair production
(Eq.~(\ref{1238})). Here we cannot explicitly
show the cancellation of neither leading nor next-to-leading logarithms
to be taken place upon combining with a corresponding virtual correction.
This is due to a partially non-analytic form of the expression for the
RC caused by hadronic VP insertion.

In Sec.~3 we examine a contribution coming from squared box type diagrams
(see Eq.~(\ref{124})) supplied by the corresponding one and two soft photon
emission contributions with the explicit expressions presented
in Eqs.~(\ref{1239},\ref{1230}).
To complete a picture we have to take into account RC caused by
genuine two-loop eikonal type amplitudes.
Keeping in mind validity of RG approach in LLA and the effect
of cancellation of large logarithms in the expression for the lowest order RC
corresponding to the eikonal type diagrams (see Ref.~\cite{BKS02})
we expect the interference between them and the Born level amplitude
to completely cancel out
upon adding to the contributions given in Eqn.~(\ref{1239}-\ref{124}).
Their explicit evaluation will be the subject of a forthcoming paper.

We are grateful for support to Heizenberg-Landau 2001-02 grant.

\section*{Appendix A}
\setcounter{equation}{0}
\renewcommand{\theequation}{A.\arabic{equation}}

In this appendix we display a set of scalar integrals to be performed
while dealing with box type diagrams bearing VP insertion into one
of the exchange virtual photon propagators.
Clearly in this case we need the integrals with a virtual exchange
photon endowed with a mass $M$. Thus in evaluating vector and tensor
integrals we use technique presented in App.~B with the only change in that
all the scalar integrals with three ($I_{ijk}$)
and four ($I$) denominators are replaced by the following ones:
\begin{itemize}
\item{in the case of large mass $M$ ($M^2\gg s\sim -t$),
\bg
I_{123}=\frac{1}{M^2}\biggl\{-\ln\frac{M^2}{s}-1
+\frac{s}{M^2}\biggl[\frac{1}{2}\ln\frac{M^2}{s}
+\frac{1}{4}\biggr]\biggr\}, \nonumber \\
I_{134}=-\frac{1}{M^2}
\biggl\{\ln\frac{M^2}{m_e^2} + 1 + \frac{t}{M^2}
\biggl[\frac{1}{2}\ln\frac{M^2}{m_e^2}+\frac{1}{4}\biggr]\biggr\}, \nonumber \\
I_{234}=-\frac{1}{M^2}\biggl\{\ln\frac{M^2}{m_\mu^2} + 1 +
\frac{t}{M^2}\biggl[\frac{1}{2}\ln\frac{M^2}{m_\mu^2}
+\frac{1}{4}\biggr]\biggr\}, \\ \nonumber
I=-\frac{1}{2sM^2}\biggl\{2\rho_s\rho_\lambda+\rho_s^2
-\frac{4\pi^2}{3}\biggr\}, \\ \nonumber
I_3=\tilde{t}I - I_{124}=\frac{1}{M^2}\biggl\{-\frac{2t}{s}\rho_s
+1-\ln\frac{s}{M^2}+\frac{s}{M^2}\biggl[\frac{1}{2}\ln\frac{s}{M^2}-\frac{1}{4}
+\left(\frac{t}{s}\right)^2\rho_s\biggr]\biggr\}\,,
\end{gather}
}
\item{
and in the opposite limit $-t\gg M^2$ we should use the next set of integrals,
\bg
I_{134}=\frac{1}{t}\biggl[\ln\frac{M^2}{m_e^2}\ln\frac{-t}{M^2}+\frac{\pi^2}{6}
+\frac{1}{2}\ln^2\frac{-t}{M^2}\biggr],  \nonumber \\
I_{234}=\frac{1}{t}\biggl[\ln\frac{M^2}{m_\mu^2}\ln\frac{-t}{M^2}+\frac{\pi^2}{6}
+\frac{1}{2}\ln^2\frac{-t}{M^2}\biggr],  \nonumber \\
I_{123}=-\frac{1}{2s}\biggl[2\rho_s\rho_m
+\frac{4\pi^2}{3} - \rho_s^2 + \ln^2\frac{m_\mu}{m_e}\biggr], \\ \nonumber
I=\frac{1}{st}\rho_s\left[\rho_\lambda + 2\rho_t - \rho_m \right], \\ \nonumber
I_{3}=\frac{1}{s}\biggl[2\rho_s\ln\frac{-t}{M^2} + \rho_s\rho_m
- \frac{1}{2}\rho_s^2 + \frac{1}{2}\ln^2\frac{m_\mu}{m_e}
+ \frac{2\pi^2}{3}\biggr].
\end{gather}
}
\end{itemize}

\section{Appendix B}
Here we put the details of box-box contribution calculation.
The first of all  we should distinguish three different cases: two box square with straight and crossed
legs and one case with interference of amplitudes with crossed and straight legs.

For calculating the contributions we need to evaluate tensor, vector and scalar integrals with
four and three denominators.

Lets first consider integral for box with straight legs. The vector integral we can write down
in the form:
\ba
\int \frac{\dd^4kk^\mu}{i\pi^2(1)(2)(3)(4)}=Ap_{1\mu}+Bp_{2\mu}+Cq_{\mu},
\ea
where quantities $(1),(2),(4)$ were defined in (\ref{eq1}), and we use here the next notation:$m=m_e$, $M=m_\mu$,
 $(3)$ is $k^2-2kq+t$,
\ba
q=p_1-p_1^{'}=p_2^{'}-p_2,\quad q^2=t,
\ea
and coefficients $A, B, C$  determined as follows:
\bg
A=\frac{1}{2stu}[-t^2a-t(2s+t)b-st c], \\ \nonumber
B=\frac{1}{2stu}[-t(2s+t)a-t^2b+stc],  \\ \nonumber
C=\frac{1}{2stu}[-sta+stb-s^2c],\\ \nonumber
a=I_{123}-I_{234},\quad b=I_{134}-I_{123},\quad c=tI.
\end{gather}
Scalar integrals $I, I_{ijk}$ reads:
\bg
\label{38}
I=\int\frac{\dd^4k}{i\pi^2(1)(2)(3)(4)}=\frac{2}{st}\biggl[\ln\frac{s}{mM}-i\pi\biggr]\ln\frac{-t}{\lambda^2},
\\ \nonumber
I_{123}=\int\frac{\dd^4k}{i\pi^2(1)(2)(3)}=-\frac{1}{2s}\biggl[2\biggl[\ln\frac{s}{mM}-i\pi\biggr]\ln\frac{\lambda^2}{mM}
+\frac{\pi^2}{3}-\biggl[\ln\frac{s}{mM}-i\pi\biggr]^2+\ln^2\frac{M}{m}\biggr], \\ \nonumber
I_{134}=\int\frac{\dd^4k}{i\pi(1)(3)(4)}=\frac{1}{t}\biggl[\frac{1}{2}\ln^2\frac{-t}{m^2}+\frac{2\pi^2}{3}\biggr], \\ \nonumber
I_{234}=\int\frac{\dd^4k}{i\pi(2)(3)(4)}=\frac{1}{t}\biggl[\frac{1}{2}\ln^2\frac{-t}{M^2}+\frac{2\pi^2}{3}\biggr].
\end{gather}
Tensor integral we consider by algebraical method:
\ba
\int\frac{\dd^4kk_{\mu}k_{\nu}}{i\pi^2(1)(2)(3)(4)}=a_gg_{\mu\nu}+a_{11}p_{1\mu}p_{1\nu}+
a_{22}p_{2\mu}p_{2\nu}+a_{12}(p_{1\mu}p_{2\nu}+p_{1\nu}p_{2\mu}) \\ \nonumber
+a_{1q}(p_{1\mu}q_{\nu}+p_{1\nu}q_{\mu})
++a_{2q}(p_{2\mu}q_{\nu}+p_{2\nu}q_{\mu})+a_{qq}q_{\mu}q_{\nu}.
\ea
To time the equation above to four-vectors $p_1, p_2, q$ we can manage the system of algebraic equations,
therefore we write down the quantities $a_{ij}$ through scalar integrals:
\bg
a_{22}=\frac{1}{s}(A_2-ta_{2q}), \\ \nonumber
a_{11}=\frac{1}{s}(A_4+ta_{1q}), \\ \nonumber
a_{12}=\frac{1}{s}(A_1-2a_g-ta_{1q}), \\ \nonumber
a_g=\frac{1}{2}(A_9-2ta_{qq}-ta_{1q}+ta_{2q}), \\ \nonumber
a_{1q}=\frac{1}{t}(A_1-A_5-ta_{2q}), \\ \nonumber
a_{2q}=\frac{1}{s+t}(A_3+A_{10}-A_5-A_9), \\ \nonumber
a_{qq}=\frac{1}{t(s+t)}(tA_3+s(A_5+A_9-A_{10})).
\end{gather}
Quantities $A_j$ read:
\bg
A_1=\frac{1}{s}(I_{13}-I_{12}), \\ \nonumber
A_2=I_{234}+\frac{1}{s}(I_{12}-I_{23})+\frac{1}{t}(2I_{34}-I_{23}-I_{24}), \\ \nonumber
A_3=I_{123}+\frac{1}{s}(2I_{12}-I_{13}-I_{23})+\frac{1}{t}(-I_{23}+I_{34}), \\ \nonumber
A_4=I_{134}+\frac{1}{s}(I_{12}-I_{13})+\frac{1}{t}(2I_{34}-I_{13}-I_{14}), \\ \nonumber
A_5=\frac{1}{s}(I_{23}-I_{12}), \\ \nonumber
A_9=tC+I_{123}+\frac{1}{s}(2I_{12}-I_{13}-I_{23}), \\ \nonumber
A_{10}=I_{123}.
\end{gather}
Here  $I_{ij}$ means scalar integrals with two denominators, $I_{ijk}$ and $I$ are determined above.
\begin{gather}
\label{42}
I_{12}=\int\frac{\dd^4k}{i\pi^2(1)(2)}=\ln\frac{\Lambda^2}{M^2}-\ln\frac{s}{M^2}+i\pi+1, \\ \nonumber
I_{13}=\int\frac{\dd^4k}{i\pi^2(1)(3)}=
I_{14}=\int\frac{\dd^4k}{i\pi^2(1)(4)}=\ln\frac{\Lambda^2}{M^2}+\ln\frac{M^2}{m^2}+1,\\ \nonumber
I_{23}=\int\frac{\dd^4k}{i\pi^2(2)(3)}=I_{24}=\int\frac{\dd^4k}{i\pi^2(2)(4)}=\ln\frac{\Lambda^2}{M^2}+1,\\ \nonumber
I_{34}=\int\frac{\dd^4k}{i\pi^2(3)(4)}=\ln\frac{\Lambda^2}{M^2}-\ln\frac{-t}{M^2}+1.
\end{gather}

For crossed legs in box type diagram we should evaluate the next integrals:
\ba
\int \frac{\dd^4kk^\mu}{i\pi^2(1)(\tilde{2})(3)(4)}=\tilde{A}p_{1\mu}-\tilde{B}p^{'}_{2\mu}+\tilde{C}q_{\mu}.
\ea
where $(\tilde{2})=k^2+p_2^{'}k$ and
\bg
\tilde{A}=\frac{1}{2stu}[-t^2\tilde{a}-t(2u+t)\tilde{b}-u\,t\, \tilde{c}], \\ \nonumber
\tilde{B}=\frac{1}{2stu}[-t(2u+t)\tilde{a}-t^2b+u\,t\,\tilde{c}],  \\ \nonumber
\tilde{C}=\frac{1}{2stu}[-u\, t\, \tilde{a}+u\, t\, \tilde{b}-u^2\, \tilde{c}],\\ \nonumber
\tilde{a}=I_{1\tilde{2}3}-I_{\tilde{2}34},\quad \tilde{b}=I_{134}-I_{1\tilde{2}3},\quad \tilde{c}=t\tilde{I}.
\end{gather}
Here integrals read:
\bg
\tilde{I}=\int\frac{\dd^4k}{i\pi^2(1)(\tilde{2})(3)(4)}=\frac{2}{ut}\ln\frac{-u}{mM}\ln\frac{-t}{\lambda^2},
\\ \nonumber
I_{1\tilde{2}3}=\int\frac{\dd^4k}{i\pi^2(1)(\tilde{2})(3)}=-\frac{1}{2u}\biggl[2\ln\frac{-u}{mM}\ln\frac{\lambda^2}{mM}
+\frac{\pi^2}{3}-\ln^2\frac{-u}{mM}+\ln^2\frac{M}{m}\biggr], \\ \nonumber
I_{\tilde{2}34}=\int\frac{\dd^4k}{i\pi(\tilde{2})(3)(4)}=\frac{1}{t}\biggl[\frac{1}{2}\ln^2\frac{-t}{M^2}+\frac{2\pi^2}{3}\biggr],
\end{gather}
and $I_{134}$ was given in (\ref{38}).

For tensor integral we have:
\ba
\int\frac{\dd^4kk_{\mu}k_{\nu}}{i\pi^2(1)(\tilde{2})(3)(4)}=\tilde{a}_gg_{\mu\nu}+\tilde{a}_{11}p_{1\mu}p_{1\nu}+
\tilde{a}_{22}p^{'}_{2\mu}p^{'}_{2\nu}-\tilde{a}_{12}(p_{1\mu}p^{'}_{2\nu}+p_{1\nu}p^{'}_{2\mu}) \\ \nonumber
+\tilde{a}_{1q}(p_{1\mu}q_{\nu}+p_{1\nu}q_{\mu})
-\tilde{a}_{2q}(p^{'}_{2\mu}q_{\nu}+p^{'}_{2\nu}q_{\mu})+\tilde{a}_{qq}q_{\mu}q_{\nu},
\ea
here we use:
\bg
\tilde{a}_{22}=\frac{1}{u}(\tilde{A}_2-t\tilde{a}_{2q}), \\ \nonumber
\tilde{a}_{11}=\frac{1}{u}(\tilde{A}_4+t\tilde{a}_{1q}), \\ \nonumber
\tilde{a}_{12}=\frac{1}{u}(\tilde{A}_1-2\tilde{a}_g-t\tilde{a}_{1q}), \\ \nonumber
\tilde{a}_g=\frac{1}{2}(\tilde{A}_9-2t\tilde{a}_{qq}-t\tilde{a}_{1q}+t\tilde{a}_{2q}), \\ \nonumber
\tilde{a}_{1q}=\frac{1}{t}(\tilde{A}_1-\tilde{A}_5-t\tilde{a}_{2q}), \\ \nonumber
\tilde{a}_{2q}=\frac{1}{u+t}(\tilde{A}_3+\tilde{A}_{10}-\tilde{A}_5-\tilde{A}_9), \\ \nonumber
\tilde{a}_{qq}=\frac{1}{t(u+t)}(t\tilde{A}_3+u(\tilde{A}_5+\tilde{A}_9-\tilde{A}_{10})).
\end{gather}
Quantities $A_j$ read:
\bg
\tilde{A}_1=\frac{1}{u}(I_{13}-I_{1\tilde{2}}), \\ \nonumber
\tilde{A}_2=I_{\tilde{2}34}+\frac{1}{u}(I_{1\tilde{2}}-I_{\tilde{2}3})+\frac{1}{t}(2I_{34}-I_{\tilde{2}3}-I_{\tilde{2}4}), \\ \nonumber
\tilde{A}_3=I_{1\tilde{2}3}+\frac{1}{u}(2I_{1\tilde{2}}-I_{13}-I_{\tilde{2}3})+\frac{1}{t}(-I_{\tilde{2}3}+I_{34}), \\ \nonumber
\tilde{A}_4=I_{134}+\frac{1}{u}(I_{1\tilde{2}}-I_{13})+\frac{1}{t}(2I_{34}-I_{13}-I_{14}), \\ \nonumber
\tilde{A}_5=\frac{1}{u}(I_{\tilde{2}3}-I_{1\tilde{2}}), \\ \nonumber
\tilde{A}_9=t\tilde{C}+I_{1\tilde{2}3}+\frac{1}{u}(2I_{1\tilde{2}}-I_{13}-I_{\tilde{2}3}), \\ \nonumber
\tilde{A}_{10}=I_{1\tilde{2}3}.
\end{gather}
Here  $I_{ij}$ means scalar integrals with two denominators, $I_{ijk}$ and $I$ are determined above.
\begin{gather}
I_{1\tilde{2}}=\int\frac{\dd^4k}{i\pi^2(1)(\tilde{2})}=\ln\frac{\Lambda^2}{M^2}-\ln\frac{-u}{M^2}+1, \\ \nonumber
I_{\tilde{2}3}=\int\frac{\dd^4k}{i\pi^2(\tilde{2})(3)}=I_{\tilde{2}4}=\int\frac{\dd^4k}{i\pi^2(\tilde{2})(4)}=\ln\frac{\Lambda^2}{M^2}+1,\\ \nonumber
\end{gather}
The other integrals $I_{13},I_{14},I_{34}$ were given in(\ref{42}).

Now with all these at hand it is straightforward to obtain
a final result for the squared box type diagrams. With the intent to
realize subsequent numeric implications let's quote it in the form
in which all the terms not enhanced by the large logarithms
are retained,
\be
\nonumber
\sum|{\cal M}_{\mathbf{box}}|^2=16\alpha^4\cdot{\cal B}(s,t,u),
\ee
\ba
{\cal B}(s,t,u)&=&\frac{8(s^2+u^2)}{t^2}(L^2_{us}+\pi^2)
\ln^2\left(\frac{-t}{\lambda^2}\right) \nonumber \\
&-& 4\ln\left(\frac{-t}{\lambda^2}\right)L_{us}\left[\frac{s-u}{t}
(L^2_{ut}+L^2_{st}-L_{ut}-L_{st}) + L_{us}\right] \nonumber \\
&+&\frac{(s-u)^2}{2}\left[\frac{1}{s^2}L^4_{ut} + \frac{1}{u^2}L^4_{st}\right]
+2(s-u)\left[-\frac{1}{s}L^3_{ut} + \frac{1}{u}L^3_{st}\right]
+ 2\left[L^2_{ut} + L^2_{st}\right]  \nonumber
\ea
\ba
\quad&+&\pi^2\Biggl\{4\ln\left(\frac{-t}{\lambda^2}\right)
\left[\frac{s-u}{t}(2L_{st}-L_{us}) +\frac{2u}{t}\right]
+\left[L_{ut}\left(1-\frac{u}{s}\right)-1\right]^2 \\ \nonumber
\quad&+& 2\left[L_{st}\left(1-\frac{s}{u}\right)-1\right]^2 - 1\Biggr\}
+\frac{\pi^4}{2}\left(1-\frac{u}{s}\right)^2.
\ea

For completeness let's present here a formula for the interference
of a tree level and a box type diagram amplitudes,
\ba
2\sum{\cal M}_{\mathbf{Born}}^\star{\cal M}_{\mathbf{box}}
&=&\sum |{\cal M}_{\mathbf{Born}}|^2\cdot\frac{\alpha}{\pi}
\Biggl\{2L_{su}(\rho_t+\rho_\lambda) \nonumber \\
&+&\frac{t^2}{s^2+u^2}\biggl[
\frac{u}{t}L_{st}-\frac{s}{t}L_{ut}
+\frac{s-u}{2t}(\pi^2+L_{ut}^2+L_{st}^2)
\biggr]\Biggr\}.
\ea
Adding to this expression the contribution arising from the ``up-down"
interference of a soft photon emission by electron and muon lines
we arrive at the expression for the RC given in Eq.~(16) of Ref.~\cite{BKS02}.

\end{document}